\documentclass[conf]{new-aiaa}
\usepackage[utf8]{inputenc}

\usepackage{amsthm}
\usepackage{amsmath}
\usepackage{amssymb}
\usepackage[version=4]{mhchem}
\usepackage{siunitx}
\usepackage{longtable,tabularx}
\usepackage{booktabs}
\usepackage{subcaption}
\usepackage{float}
\usepackage{graphicx}
\usepackage[svgnames]{xcolor}
\theoremstyle{definition}
\newtheorem{assumption}{Assumption}

\newcommand{\ra}[1]{\renewcommand{\arraystretch}{#1}}
\newcommand{\eqdef}{\triangleq}

\DeclareMathOperator*{\argmin}{\arg\!\min}

\newcommand{\btheta}{\boldsymbol\theta}
\newcommand{\thetadot}{\dot{\theta}}
\newcommand{\Phidot}{\dot{\Phi}}
\newcommand{\Q}{\mathbf{Q}}

\newcommand{\yref}{y_\text{r}}
\newcommand{\Cd}{C_\text{d}}
\newcommand{\Chatd}{\hat{C}_\text{d}}
\newcommand{\Ctilded}{\tilde{C}_\text{d}}
\newcommand{\Cdmax}{\bar{C}_\text{d}}
\newcommand{\Lcm}{L_\text{cm}}

\usepackage{changes}
\definechangesauthor[name={B.~L.}, color={blue}]{bl}
\definechangesauthor[name={J.~H.}, color={red}]{jh}
\definechangesauthor[name={P.~L.}, color={orange}]{pl}
\definechangesauthor[name={S.~M.}, color={purple}]{sm}

\title{Performance Analysis of Adaptive Dynamic Tube MPC}

\author{Savva Morozov\footnote{Undergraduate student, MIT Aerospace Controls Laboratory} and Parker C.\ Lusk\footnote{Graduate research assistant, MIT Aerospace Controls Laboratory}}
\affil{Massachusetts Institute of Technology, Cambridge, MA, 02141}
\author{Brett T.\ Lopez\footnote{Postdoctoral researcher, NASA Jet Propulsion Laboratory}}
\affil{NASA Jet Propulsion Laboratory, Pasadena, CA, 91109}
\author{Jonathan P.\ How\footnote{Richard C.\ Maclaurin Professor, Department of Aeronautics and Astronautics, MIT Aerospace Controls Laboratory.}}
\affil{Massachusetts Institute of Technology, Cambridge, MA, 02141}

\begin{document}

\maketitle

\begin{abstract}

Model predictive control (MPC) is an effective method for control of constrained systems but is susceptible to the external disturbances and modeling error often encountered in real-world applications.
To address these issues, techniques such as Tube MPC (TMPC) utilize an ancillary offline-generated robust controller to ensure that the system remains within an invariant set, referred to as a tube, around an online-generated trajectory.
However, TMPC is unable to modify its tube and ancillary controller in response to changing state-dependent uncertainty, often resulting in overly-conservative solutions.
Dynamic Tube MPC (DTMPC) addresses these problems by simultaneously optimizing the desired trajectory and tube geometry online.
Building upon this framework, Adaptive DTMPC (ADTMPC) produces better model approximations by reducing model uncertainty, resulting in more accurate control policies.
This work presents an experimental analysis and performance evaluation of TMPC, DTMPC, and ADTMPC for an uncertain nonlinear system. 
In particular, DTMPC is shown to outperform TMPC because it is able to dynamically adjust to changing environments, limiting aggressive control and conservative behavior to only the cases when the constraints and uncertainty require it.
Applied to a pendulum testbed, this enables DTMPC to use up to 30\% less control effort while achieving up to 80\% higher speeds.
This performance is further improved by ADTMPC, which reduces the feedback control effort by up to another 35\%, while delivering up to 34\% better trajectory tracking.
This analysis establishes that the DTMPC and ADTMPC frameworks yield significantly more effective robust control policies for systems with changing uncertainty, goals, and operating conditions.

\end{abstract}

\section{Introduction}\label{sec:introduction}

Model predictive control (MPC) has become a foundational strategy for many applications including trajectory generation~\cite{singh2001trajectory}, collision avoidance~\cite{ji2016path, wang2007cooperative}, and sequential decision making processes~\cite{russell2002artificial}.
MPC is widely applicable due to its ability to handle constraints while balancing competing objectives.
However, high reliance on a dynamical model makes MPC susceptible to modeling error and external disturbances.
These uncertainties can therefore substantially reduce the effectiveness of MPC and can even cause instability if not accounted for~\cite{mayne2000constrained}. 
The purpose of this work is to analyze the hardware performance of MPC algorithms that consider these uncertainties, emphasizing our new DTMPC and ADTMPC frameworks, which account for state-dependent uncertainty in trajectory generation and control.

One of the first such methods is Robust MPC (RMPC)~\cite{bemporad1999robust}, which searches over a high-dimensional space of feedback control policies for the one that provides robust tracking under known uncertainty, yet this method is computationally expensive and is often intractable~\cite{althoff2008reachability}. 
Tube MPC (TMPC)~\cite{langson2004robust} addresses these issues by using an ancillary robust controller, designed offline, to maintain the system within an invariant tube around an online-generated desired trajectory.
This decoupled trajectory-control design causes TMPC to utilize a constant-size tube regardless of the system's state-dependent uncertainty, often producing conservative results under changing objectives and environments.
Attempts to vary the tube size throughout the trajectory have been made, such as Homothetic and Elastic Tube MPCs~\cite{rakovic2012homothetic, rakovic2016elastic}, but these algorithms do not explicitly consider model uncertainty and only apply to linear systems.
A wealth of nonlinear Tube MPC approaches have been proposed, utilizing reachability theory~\cite{althoff2008reachability}, sliding mode control~\cite{rubagotti2010robust}, and control contraction metrics~\cite{singh2017robust}.
Although robust, these methods tend to produce high-bandwidth controllers, which can negatively impact actuator lifespan and are often impractical if on-board state-estimation and perception are used~\cite{faessler2015automatic}.
Additionally, these methods lack the ability to modify the tube online.

Attempts to vary the tube size in nonlinear systems include the work by Lakshmanan et al.~\cite{lakshmanan2020safe}, which uses contraction theory and L1-adaptive control to parameterize the tube. 
However, it is unclear how to use this parameterization to vary the tube size in response to uncertainty.
Fan et al.~\cite{fan2020deep} proposed using quantile regression with deep learning to learn the uncertainty-dependent dynamics of a probabilistic tube around the trajectory.
Yet this method only quantifies the performance of existing policies and does not provide an intuitive way to tune that policy based on desired tube behavior.

To address the lack of a nonlinear control strategy that could simultaneously optimize the desired trajectory and the tube geometry in response to changing uncertainty, we recently presented Dynamic Tube MPC (DTMPC)~\cite{lopez2019dynamic}.
We derived a direct relationship between model uncertainty and the tube's radius through a control bandwidth variable that is shown to govern feedback aggressiveness.
By using a nonlinear ancillary controller that admits such a relationship, DTMPC is able to respond to changing state-dependent constraints and uncertainty online in a computationally efficient manner that is also applicable to nonlinear systems.

Adaptive Dynamic Tube MPC (ADTMPC) builds upon the DTMPC controller by introducing an adaptation framework~\cite{lopez2019thesis}.
It uses state membership identification (SMID) to eliminate uncertain parameter values that are not consistent with the dynamics, the disturbance model, and the observations.
This procedure is derived in such a way as to maintain controller robustness throughout online adaptation, generating consistently feasible, less-conservative behaviors as a result of acquiring more accurate model predictions.

The main contribution of this work is the performance analysis of DTMPC and ADTMPC algorithms as applied to a physical nonlinear pendulum.
We first draw the comparison between DTMPC and its predecessor, TMPC.
We show that DTMPC's ability to leverage changing constraints and uncertainty results in noticeable reductions in utilized control effort and trajectory conservativeness.
This makes DTMPC a more suitable controller for operation under changing environments and objectives.
We then illustrate how ADTPMC further improves on DTMPC's performance.
We show ADTMPC's ability to estimate the true values of unknown parameters, producing steadily more accurate and less conservative control policies as a result of adaptation. 
Finally, we demonstrate that ADTMPC achieves consistent results under different hardware configurations, control settings, and trajectories.

The structure of this paper is as follows.
Section~\ref{sec:prelim} introduces mathematical preliminaries which lead up to the presentation of DTMPC and ADTMPC.
In Section~\ref{sec:probform}, we apply DTMPC and ADTMPC to our pendulum system. 
We then introduce our hardware testbed in Section~\ref{sec:implementation}, define the experiments, and present the results in Sections~\ref{sec:results-dtmpc} and~\ref{sec:results-adtmpc}, followed by a discussion in Section~\ref{sec:discussion}.
We conclude with final thoughts and future work in Section~\ref{sec:conclusion}.

\section{Preliminaries}\label{sec:prelim}

In this paper we are concerned with tracking trajectories of feedback linearizable nonlinear systems in the form of 
\begin{equation}\label{eq:nlsys}
    \dot{x} = f(x) + b(x) u + d,\quad\; y = h(x),
\end{equation}
where $x \in\mathbb{X}\subseteq\mathbb{R}^n$ is the state, $u\in\mathbb{U}\subseteq\mathbb{R}^m$ is the control input, $\mathbb{X}$ and $\mathbb{U}$ represent the constraints on state and input respectively, $d \in\mathbb{R}^n$ is an external disturbance, and $y\in\mathbb{R}^m$ is the system's output.
Although these variables are functions of time, we will frequently drop time dependence for presentation clarity.

\begin{assumption}
\label{assum:1}
The external disturbance $d$ is unknown but bounded, i.e., $|d(t)|\le D$.
Additionally, we assume that $d(t)$ lies in the span of the control input matrix, $d\in \text{range}\big(b(x)\big).$ 
\end{assumption}

\begin{assumption}
\label{assum:2}
The dynamics $f$ can be expressed as $f=\hat{f}+\tilde{f}$ where $\hat{f}$ is the nominal dynamics and $\tilde{f}$ is the bounded model error, i.e., $|\tilde{f}|\le\Delta(x)$.
\end{assumption}

\begin{assumption}
\label{assum:3}
The input matrix $b(x)$ is known.
This assumption can be easily relaxed as in~\cite{lopez2019thesis}.
\end{assumption}

Since the nonlinear system is feedback linearazible, it can be transformed into an equivalent linear system whose states are the initial system's output and its first $n-1$ derivatives.
Let the desired trajectory be defined as $x^*(t)$, together with its corresponding feedback-linearized trajectory $y^*(t)$ and corresponding open-loop control input $u^*(t)$.
Tracking error is defined as $\tilde{y}(t)\eqdef y(t)-y^*(t)$.
In the following subsections, we recall boundary layer sliding control and review the differences between traditional MPC, TMPC, DTMPC, and ADTMPC.

\subsection{Boundary Layer Sliding Control}\label{sec:blsc}

Sliding mode control (SMC) is a robust nonlinear trajectory tracking control strategy applicable to feedback linearizable systems as in~\eqref{eq:nlsys}.
SMC operates by simplifying an $n^\text{th}$-order tracking problem into $m$ first-order stabilization problems.
To this end, we define a scalar time-varying surface $S(t) = \{y(t) : s(y;t)=0\}$, where $s(y;t)$ is called the sliding variable, defined as 
\begin{align}\label{eq:sliding-variable}
    s(y;t) &= \left(\frac{d}{dt} + \lambda\right)^{n-1} \tilde{y}(t) \\
    &= \tilde{y}^{(n-1)} + \cdots + \lambda^{n-1}\tilde{y} = y^{(n-1)} - \yref^{(n-1)}.
\end{align}

Note that $\yref^{(n-1)}$ is a notational convenience for the so-called ``reference'' value of $y^{(n-1)}$.
Once $s(y;t)=0$, the tracked outputs of the dynamical system~\eqref{eq:nlsys} lie on the surface $S$. 
The tracking error then decays exponentially to zero with the time constant $\lambda>0$ according to~\eqref{eq:sliding-variable}, even in the presence of bounded modeling error or bounded external disturbance.
This robustness is often obtained via a discontinuous control law and comes at the cost of chattering, a well-known high-frequency switching phenomenon typically unsuitable for many real-world systems.
Chattering can be eliminated with the use of a deadband, or boundary layer, with thickness $\Phi$.
For even better performance, the boundary layer thickness may be time-varying ($\dot{\Phi}\neq0$) so that it may shrink and grow according to the current uncertainty affecting the system.
These reformulations constitute the boundary layer sliding control (BLSC) and were originally developed in~\cite{slotine1984sliding}.
The continuous control law that leads the system to the sliding surface $S(t)$ can be written as
\begin{align}
u &= u^* + \tilde{u} = u^* + b(x)^{-1} \left[-\hat{f}(x) - y_r^{(n)} - K(x)\;\mathrm{sat}\left(s/\Phi\right)\right], \label{eq:blsc-u} \\
K(x) &= \Delta(x) + D + \eta - \dot{\Phi} = \alpha \Phi, \label{eq:k-def} \\
 \dot{s} &= -\alpha s  + \tilde{f}(x) + d, \label{eq:blsc-sdot} \\
 \dot{\Phi} &= -\alpha\Phi + \Delta(x^*) + D + \eta, \label{eq:blsc-phidot}
\end{align}
where $\eta>0$ is the convergence rate towards the sliding surface. 
Given the control law in~\eqref{eq:blsc-u} and~\eqref{eq:k-def}, we obtain the dynamics of the sliding variable $s$ and its boundary layer $\Phi$, described in~\eqref{eq:blsc-sdot} and~\eqref{eq:blsc-phidot}.
In these two equations, the time constant $\alpha>0$ can be viewed as the bandwidth of first-order filters on the sliding variable and its boundary layer, respectively driven by state-dependent uncertainty $\tilde{f}$ and by its upper boundary $\Delta$.
This observation suggests that $\alpha$ is also the bandwidth of a low-pass filter on the last term of the ancillary control law in~\eqref{eq:blsc-u}, the term that is dominant when state-dependent uncertainty is present.
This relationship is why $\alpha$ is referred to as the control bandwidth of this ancillary BLSC controller: by varying $\alpha$ we can vary the aggressiveness of the stabilizing controller. 
We refer the reader to~\cite{slotine1984sliding,slotine1991applied,lopez2019thesis} for further details on the derivation of BLSC.

\subsection{Traditional MPC}\label{sec:mpc}

MPC involves repeatedly solving a nonlinear optimal control problem over a time horizon while directly accounting for state and actuator constraints.
The incorporation of constraints ensures that the system complies with its hardware limitations, which is the main benefit compared to classical control methods.
A typical MPC problem can be formulated as the following optimization.

\vspace{1em}
\noindent\textbf{Problem 1} --- Traditional MPC
\begin{equation*}
\begin{aligned}
u^*(t) \; = \;\; & \underset{ \check{x}(t), \check{u}(t) }{\argmin} & & h(\check{x}(t_f))  + \int_{t_0}^{t_f} l( \check{x}(t), \check{u}(t) )  dt   \\
& \text{subject to} & & \dot{\check{x}}(t) = \hat{f}(\check{x}(t)) + b(\check{x}(t)) \, \check{u}(t), \\
&&&  \check{x}(t_0) = x(t_0), \quad \check{x}(t_f) \in \mathbb{X}_f, \\
&&& \check{x}(t) \in \mathbb{X}, \quad \check{u}(t) \in \mathbb{U},
\end{aligned}
\end{equation*}
where $h(\cdot)$ is a terminal state quadratic cost, $l(\cdot)$ is the combined state and actuator stage cost, $u^*$ is the resulting open-loop optimal control policy, $\check{\cdot}$ denotes internal optimization variables,
$x(t_0)$ is the current state of the system,
and $\mathbb{X}_f$ is the set of terminal states, such that $\mathbb{X}_f \subseteq \mathbb{X}$.
Although Problem~1 can be solved online, MPC has to execute open-loop inputs until re-solving, making it highly dependent on the model which, if wrong, can lead to poor performance~\cite{mayne2000constrained}.

\subsection{Tube MPC}\label{sec:tmpc}

\begin{figure}[t]
    \centering
    \includegraphics[width=0.5\textwidth]{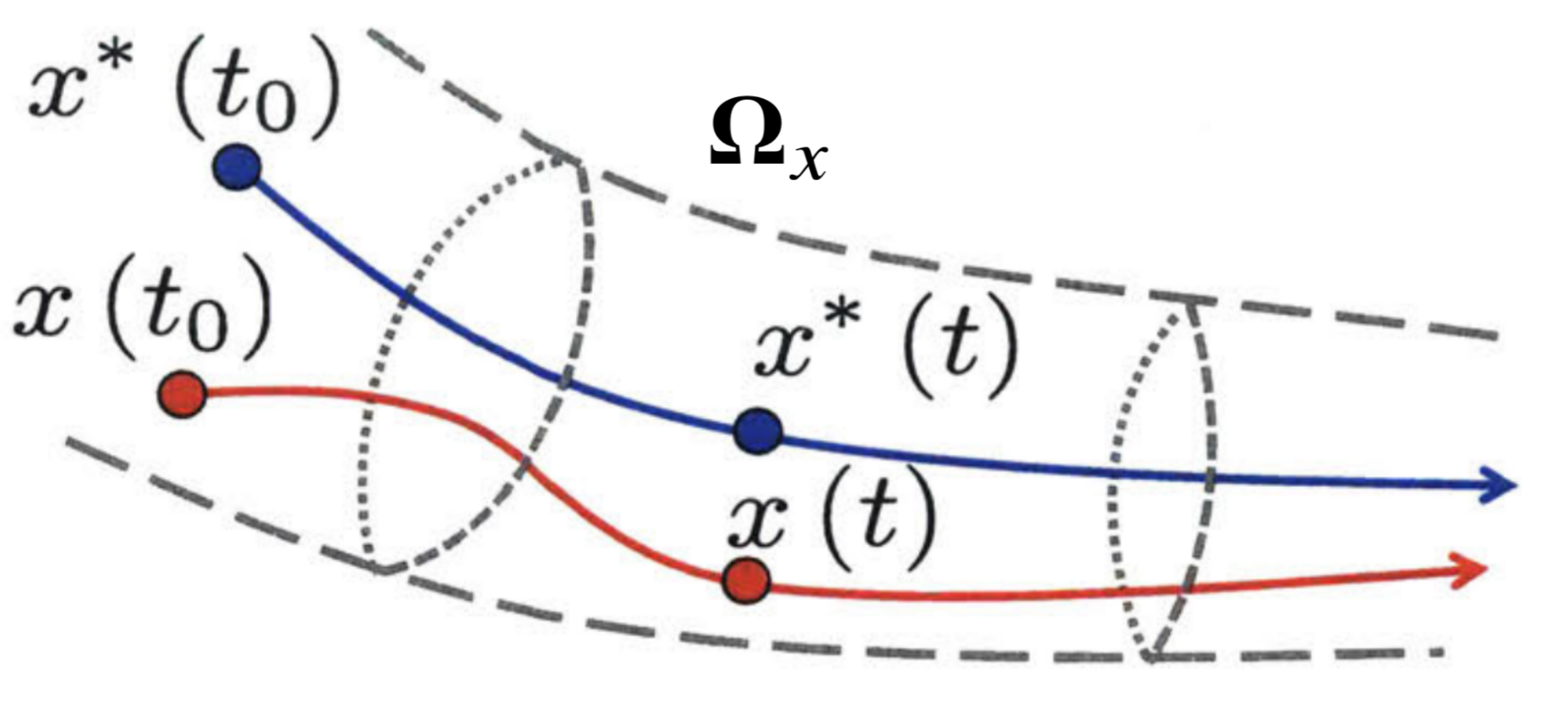}
    \caption{
    The robust control invariant (RCI) tube $\boldsymbol{\Omega}_{x}$ around the desired trajectory $x^*$.
    If the state $x$ begins inside the RCI tube, it remains inside the RCI tube indefinitely for all realizations of uncertainty~\cite{mayne2011tube}. 
    }
    \label{fig:rci}
\end{figure}

TMPC separates the control problem into two subproblems: trajectory generation and trajectory tracking.
This decoupling assumes that the control policy is of the form $\pi = u^* + \kappa(x,x^*)$, where $u^*$, $x^*$ are the open-loop control and the corresponding desired state as generated by MPC, and $\kappa(x,x^*)$ is the ancillary feedback-control law designed offline.
The ancillary controller is designed to maintain the system within a \textit{robust control invariant} (RCI) tube centered around the nominal trajectory $x^*(t)$, as shown in Fig.~\ref{fig:rci}.
The RCI tube is the set $\boldsymbol\Omega_{x}$ of allowed tracking errors $\tilde{x}$, such that for all disturbance and modeling error realizations, $\tilde{x} \in \boldsymbol\Omega_{x}$.

To prevent the ancillary controller from causing $\pi$ to violate state and actuator constraints, \textit{constraint tightening} is performed wherein $\mathbb{X}$ and $\mathbb{U}$ are modified to account for the feedback control and for the geometry of the RCI tube.
Constraint tightening is a necessary procedure when an ancillary controller is utilized.
A typical formulation of the TMPC control policy takes the following form.

\vspace{1em}
\textbf{Problem 2} --- Tube MPC
\begin{equation*}
\begin{aligned}
x^*(t), \, u^*(t) \; = \;\; & \underset{ \check{x}(t), \check{u}(t) }{\argmin} & & h(\check{x}(t_f))  + \int_{t_0}^{t_f} l( \check{x}(t), \check{u}(t) )  dt   \\
& \text{subject to} & & \dot{\check{x}}(t) = \hat{f}(\check{x}(t)) + b(\check{x}(t)) \, \check{u}(t), \\
&&&  \check{x}(t_0) = x_0^*, \quad \check{x}(t_f) \in \overline{\mathbb{X}}_f, \\
&&& \check{x}(t) \in \overline{\mathbb{X}}, \quad \check{u}(t) \in \overline{\mathbb{U}}
\end{aligned}
\end{equation*}
where $\overline{\mathbb{X}}$, $\overline{\mathbb{X}}_f$, and $\overline{\mathbb{U}}$ represent tightened state, terminal state, and actuator constraints, respectively.
Note that the initial state is usually chosen not to be the current state $x(t_0)$ but some state along a previously generated trajectory, as the open loop trajectories are robustly tracked.
Since the above formulation is just a modified version of Problem~1, TMPC adds no extra online complexity to solving the MPC problem.
Rather, the complexity of the algorithm is in designing an ancillary controller $\kappa(x, x^*)$ together with the RCI tube $\boldsymbol\Omega_{x}$, both of which are done offline.
As an example, one choice of ancillary controller is the family of sliding mode controllers~\cite{rubagotti2010robust,muske2007predictive,incremona2015hierarchical}.

Note that by decoupling the trajectory and ancillary controller design, TMPC is inherently limited as the nominal MPC optimization is unable to modify the ancillary controller in the presence of changing uncertainty.
We stress that when the RCI tube is generated offline, the worst-case uncertainty is assumed, resulting in conservative tube geometry.
This often leads to reduced performance and can even result in trajectory infeasibility.
These challenges are overcome with DTMPC, as described in the next section.

\subsection{Dynamic Tube MPC}\label{sec:dtmpc}

DTMPC utilizes time-varying BLSC as its ancillary controller. 
In equation~\eqref{eq:blsc-phidot} we established the relationship between
uncertainty
and the boundary layer thickness $\Phi$. 
DTMPC then defines a relationship between $\Phi$ and the radius $\Omega$ of the RCI tube around the nominal trajectory.
To respond to changing uncertainty, DTMPC makes the control bandwidth $\alpha$ a time-varying optimization decision variable, which we elaborate on below.
This framework successfully couples trajectory generation with tube geometry design and incorporates state-dependent uncertainty into the optimization~\cite[Thereom 1]{lopez2019dynamic}.
The general formulation of DTMPC is presented below.

\vspace{1em}
\textbf{Problem 3} --- Dynamic Tube MPC
\begin{equation*}
\begin{aligned}
x^*(t), \, u^*(t), \, \alpha^*(t) \; = \;\; & \underset{ \check{x}(t), \, \check{u}(t), \, \check{\alpha}(t) }{\argmin} & & h(\check{x}(t_f))  + \int_{t_0}^{t_f} l( \check{x}(t), \check{u}(t), \check{\alpha}(t) )  dt   \\
& \text{subject to} & & \dot{\check{x}}(t) = \hat{f}(\check{x}(t)) + b(\check{x}(t)) \, \check{u}(t), \\
& & & \dot{\Phi}(t) = -\check{\alpha}(t) \Phi(t) + \Big( \Delta(\check{x}(t)) + D + \eta \Big),  \\
& & & \dot\Omega(t) = A_\text{c} \Omega(t) + B_\text{c} \Phi(t), \quad \Omega(t_0) = |\tilde{x}(t_0)|, \\
& & & \check{x}(t_0) = x_0^*, \quad \Phi(t_0) = \Phi_0, \\
& & & \check{x}(t) \in \overline{\mathbb{X}}, \quad \check{u}(t) \in \overline{\mathbb{U}}, \quad \check{\alpha}(t) \in \mathbb{A}, \quad \dot{\check{\alpha}}(t) \in \mathbb{V}
\end{aligned}
\end{equation*}
where $A_\text{c}$, $B_\text{c}$ matrices are derived such that $\Omega$ defines an RCI Tube (we refer the reader to \cite{lopez2019dynamic} for exact matrix derivation).
The bandwidth $\alpha$ determines the response of $\Phi$ to changing uncertainty.
Note that uncertainty drives the boundary layer thickness $\Phi$, which in turn drives the tube radius $\Omega$. 
Therefore $\alpha$ indirectly determines the response of $\Omega$ to changing uncertainty, making $\alpha$ a logical choice for the decision variable. 

The sets $\overline{\mathbb{X}}$ and $\overline{\mathbb{U}}$ represent tightened state and actuator constraints that take into account state-dependent uncertainty. 
As DTMPC leverages this uncertainty by varying the size of the tube, it also varies the state and actuator constraints throughout the trajectory.
This is done to ensure that the system does not exceed its physical limitations when dealing with specific uncertainty realizations.
Lastly, $\mathbb{A}$ and $\mathbb{V}$ represent the constraints on control bandwidth and its derivative.

The benefit of DTMPC comes from capturing the relationship between state-dependent uncertainty and RCI tube geometry. 
As a consequence of this relationship, DTMPC only displays aggressive behavior when it absolutely needs to---when the system is close to its constraints.
In all other conditions, DTMPC will reduce expended control effort as compared to its TMPC predecessor.
Likewise, DTMPC only displays conservative behaviors when uncertainty is high, tightening the constraints online in response to changing uncertainty and producing less conservative solutions overall.

While DTMPC does increase computational complexity of the problem by introducing additional constraints and an extra differential equation, this additional complexity is not significant~\cite{lopez2019dynamic}. 
Since uncertainty is directly incorporated into the model, DTMPC is also beneficial in that it is one of the only algorithms that allows improving a nonlinear model online, making it possible to implement adaptive control methods that could reduce model uncertainty.

\subsection{Adaptive Dynamic Tube MPC}

Adaptive DTMPC uses state membership identification (SMID) to reduce the model uncertainty $\tilde{f}$ online.
The following assumption is made about uncertain model dynamics:

\begin{assumption}
\label{assum:4}
The dynamics $f$ can be expressed as a linear function of a parameter vector $\rho$, which is the sum of the nominal value $\hat{\rho}$ and the error term $\tilde{\rho}$: $f = \phi(x) \rho =  \phi(x) (\hat{\rho} + \tilde{\rho}).$ Additionally, the true value of the parameter vector $\rho$ must belong to a known, closed, convex set $P$.
\end{assumption}

Rewriting~\eqref{eq:nlsys} and using Assumptions \ref{assum:1} and \ref{assum:4}, we get: 
\begin{align}
& | \dot{x} - \phi(x) \rho - b(x) u| = |d| \leq D. \label{eq:smid-gen}
\end{align}

Notice that~\eqref{eq:smid-gen} defines a constrained convex set $\Xi$ that further constrains the parameter vector $\rho$. 
After obtaining measurements $x$, $\dot{x}$, and $u$, we calculate the set $\Xi$ and update the parameter set $P$ via the update rule
\begin{align}
& \Xi = \{ \rho \in P : | \dot{x} - \phi(x) \rho - b(x) u| \leq D \} \label{eq:smid-xhi} \\
& P = P \cap \Xi. \label{eq:smid-upd}
\end{align}

Running ADTMPC online requires \textit{recursively feasibility}, meaning that the problem remains feasible for all time~\cite{lofberg2012oops}.
This demands that the system is maintained in the RCI tube for any $\rho\in P$, and for the uncertainty to be reduced monotonically throughout adaptation.
The first requirement is made possible by the BLSC ancillary controller, which is capable of not only stabilizing the system, but also maintaining it within an RCI tube for all bounded parameter uncertainties and unmodeled disturbances.
The second is satisfied by equation~\eqref{eq:smid-upd}, which ensures that the parameter set $P$ decreases monotonically.
A full proof of monotonicity and the recursive feasibility of ADTMPC can be found in~\cite{lopez2019thesis}. 

SMID is not computationally expensive, as it only entails solving linear programs, and therefore will not hinder the overall performance. 
Additionally, this procedure needs not be executed at the same rate as the BLSC or DTMPC controllers, but can instead be run in parallel and at a slower rate.
We note that unlike other adaptive methods, SMID does not require excitation in the model dynamics, but instead requires excitation in the disturbances and unmodeled dynamics $d$.
Therefore a tighter approximation of $D$ would produce better results.

\section{Problem Formulation}\label{sec:probform}

\begin{figure}[t]
    \centering
    \includegraphics[width=0.4\textwidth]{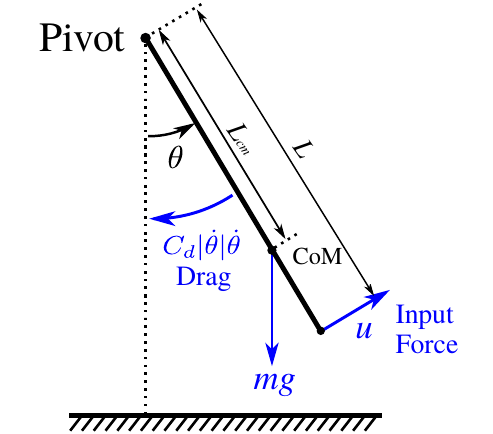}
    \caption{
    A free body diagram of a pendulum.
    }
    \label{fig:free-body}
\end{figure}

To analyze algorithmic performance, we use a pendulum testbed as depicted in Fig.~\ref{fig:free-body}.
This choice provides sufficient complexity, nonlinearity, and parameter uncertainty to be able to leverage DTMPC and ADTMPC.
We consider the following uncertain dynamics of the pendulum
\begin{equation} \label{eq:pend}
\ddot{\theta} + \frac{\Cd}{I} |\dot{\theta}|\dot{\theta} + \frac{\Lcm mg}{I} \sin(\theta) = \frac{L}{I}u + d
\end{equation}
where $\theta$ is the arm angle as measured from the arm-down position, $I$ is the inertia around the axis of rotation, $L$ is the distance from the axis at which input force is applied, $\Lcm$ is the distance from the axis to the pendulum's center of mass, $m$ is the  mass of the pendulum, and $g$ is the gravitational acceleration.
The coefficient $\Cd$ is an unknown but bounded angular drag coefficient ($0\le\Cd\le\Cdmax=\Chatd+\Ctilded$),
 and $d$ captures all other bounded disturbances ($|d|\le D$).

Given these dynamics, we define the sliding variable $s=\dot{\tilde{\theta}}+\lambda\tilde{\theta}$ and apply the BLSC methodology defined in Section~\ref{sec:blsc} to obtain
\begin{align}
\label{eq:pend_input}
u &= u^* + \frac{I}{L} \left[
    \frac{L_\text{cm} m g}{I} \left(\sin\theta - \sin\theta^*\right)
    + \frac{\Chatd}{I} \left(
        |\thetadot|\thetadot - |\thetadot^*|\thetadot^*
    \right)
    - \lambda \dot{\tilde{\theta}}
    - K \mathrm{sat}\left(\frac{s}{\Phi}\right)
\right] \\
K &= \Delta(\thetadot^*) + D + \eta - \Phidot = \alpha^* \Phi \label{eq:pend_K} \\
\Phidot &= -\alpha^* \Phi + \Delta(\thetadot^*) + D + \eta, \label{eq:pend_phidot}
\end{align}
where $\Chatd$ is the best estimate of the drag coefficient, and $\Delta(\thetadot)\eqdef\frac{\Ctilded}{I}|\thetadot|\thetadot$.

As discussed in Section~\ref{sec:tmpc}, the use of a tube requires a constraint tightening step to account for uncertainty along the open-loop trajectory.
In particular, since DTMPC produces a time-varying tube, the constraints are also time-varying and are modified during the optimization of the trajectory.
To find these tightened constraint bounds, we first define $\dot{\tilde{\theta}}_\text{max}$, the time-varying upper bound on the tracking error of angular speed $\dot{\tilde{\theta}}$,

\begin{equation}\label{eq:sc1}
|\dot{\tilde{\theta}}|
= |s - \lambda \tilde{\theta}| \le \Phi + \lambda \Omega
\eqdef \dot{\tilde{\theta}}_\text{max}(t).
\end{equation}
Using $\dot{\tilde{\theta}}_\text{max}$, we calculate the tightened, time-varying state-dependent constraints on state and input to be
\begin{align}
|\dot{\theta}^*| &\le \dot{\theta}_\text{max} - \dot{\tilde{\theta}}_\text{max}  \label{eq:sc2} \\
|u^*| &\le u_\text{max}  - \frac{I}{L}\left( \frac{L_\text{cm} m g}{I} \left(2 \sin(\Omega/2) \right) + \frac{\hat{C}_\text{d}}{I} \dot{\tilde{\theta}}_\text{max} (2 \dot{\theta}_\text{max}) + \lambda \dot{\tilde{\theta}}_\text{max} + \alpha^* \Phi \right) = u_\text{max} - \tilde{u}_\text{max}, \label{eq:sc3}
\end{align}
where $\tilde{u}_\text{max}$ represents the upper bound on the ancillary BLSC input required to maintain trajectories of the system within the RCI tube.
We are now able to give a complete formulation of the pendulum DTMPC.

\vspace{1em}
\noindent\textbf{Problem 3.1} --- Pendulum DTMPC
\begin{equation*}
\begin{aligned}
& \underset{ \check{\btheta}(t),  \check{u}(t),\check{\alpha}(t) }{\argmin}
& & (\check{\btheta}(t_f) - \btheta^*_f)^\top \Q_f (\check{\btheta}(t_f) - \btheta^*_f) \\
& & & \qquad   + \int_{t_0}^{t_f} \left(  
        (\check{\btheta}(\tau) - \btheta^*_f)^\top \Q (\check{\btheta}(\tau) - \btheta^*_f)
        + R(\check{u}(\tau) - u^*_f)^2 
        + M(\check{\alpha}(\tau) - \alpha_\text{min})^2 
    \right) d\tau \\
& \text{subject to}
& & \ddot{\check{\theta}}(t) = \frac{L}{I} \check{u}(t) - \frac{\Chatd}{I} |\dot{\check{\theta}}(t)| \dot{\check{\theta}}(t) - \frac{\Lcm mg}{I} \sin(\check{\theta}(t)), \\
& & & \Phidot(t) = -\check{\alpha}(t) \Phi(t) + \left( \Delta(\dot{\check{\theta}}(t)) + D + \eta \right),  \\
& & & \dot\Omega(t) = - \lambda \Omega(t) + \Phi(t),
        \quad  0 \le \Omega(t) \le \Omega_{\text{max}}, \\
& & & \Omega(t_0) = |\tilde{\theta}(t_0)|,
        \quad \check{\btheta}(t_0) = \btheta^*_0,
        \quad \Phi(t_0) = \Phi_0, \\
& & & | \dot{\check{\theta}}(t)| \le \thetadot_\text{max} - \dot{\tilde{\theta}}_\text{max}(t),
        \quad | \check{u}(t)| \le u_\text{max} - \tilde{u}_\text{max}(t), \\
& & & \alpha_\text{min} \le  \check{\alpha}(t) \le \alpha_\text{max},
        \quad |\dot{\check{\alpha}}(t)| \le \Delta \alpha_\text{max},
        \quad  |\dot{\check{u}}(t)| \le \Delta u_\text{max}
\end{aligned}
\end{equation*}
where $\btheta=\left[\theta\hspace{0.75em}\thetadot\right]^\top$, quantities $\Q_f$, $\Q$, $R$, $M$ are penalties for terminal state, state, input, and control bandwidth respectively,  $\boldsymbol{\theta}^*_f$ and $u^*_f$ are the desired terminal state and input, $\alpha_\text{min}$ and $\alpha_\text{max}$ are the lower and upper boundaries on control bandwidth, $\Omega_{\text{max}}$ is the upper bound on the RCI tube radius, and $\Delta u_\text{max}$, $\Delta \alpha_\text{max}$ are the constraints on the derivatives of input and control bandwidth that ensure that generated trajectories are smooth and match physical hardware capabilities.

ADTMPC extends DTMPC with the SMID framework that is ran in parallel.
Suppose $\rho$ is one of the four parameters \{$I$, $L$, $C_d$, $L_{\text{cm}}$\} of our system, and $\mathbf{P}$ is a convex parameter set such that $\rho \in \mathbf{P}$.
After acquiring state measurements $\theta$, $\dot\theta$, $\ddot\theta$ and control input $u$, we find new lower and upper boundaries on the set $\mathbf{P}$ by solving the following linear program (LP)

\vspace{1em}
\textbf{Problem 3.2}---Pendulum ADTMPC: State Membership Identification
\begin{equation*}
\begin{aligned}
q\, = \,\underset{\rho}{\text{argmin}}\; c \rho &\\
\text{subject to} &\quad |\,  I \ddot{{\theta}} - L u + C_d |\dot{{\theta}}| \dot{{\theta}} + L_{\text{cm}} mg \sin({\theta})  \,|\, \leq\, D I \\
&\quad I \in \mathbf{I}, \quad L \in \mathbf{L}, \quad C_d \in \mathbf{C}_d, \quad L_{\text{cm}} \in \mathbf{L}_{\text{cm}}\\
\end{aligned}
\end{equation*}
where $q$ is the new lower or upper boundary on the parameter $\rho$, 
and $c$ is a coefficient that is either positive or negative, which determines whether the lower or the upper boundary on $\rho$ is calculated.
By solving eight such LPs, we can update each of the parameter sets $\mathbf{P}$ $\in$ \{$\mathbf{I}$, $\mathbf{L}$, $\mathbf{C}_d$, $\mathbf{L}_{\text{cm}}$\}.
Although uncertainties on $I$, $L$, and $L_{\text{cm}}$ exist, they are negligible and hence do not appear in the DTMPC formulation.
This simplification is justified, since we can directly measure these three parameters.
However, within the SMID framework we account for these parametric uncertainties to ensure algorithmic convergence.

Lastly, we formulate pendulum TMPC. 
To make it easier to draw parallels between DTMPC and TMPC, as well as to ensure a fair comparison between the controllers, our implementation of TMPC will also use BLSC for its ancillary policy.
Since TMPC is defined to have a rigid constant-radius tube, this BLSC must have constant-thickness boundary layer, $\dot{\Phi} = 0$. 
We are hence able to define TMPC as a simplified case of the above DTMPC formulation by setting the tube to be constant.

\vspace{1em}
\textbf{Problem 3.3}---Pendulum TMPC
\begin{equation*}
\begin{aligned}
& \underset{ \check{\btheta}(t),  \check{u}(t)}{\argmin}
& & (\check{\btheta}(t_f) - \btheta^*_f)^\top \Q_f (\check{\btheta}(t_f) - \btheta^*_f)
    + \int_{t_0}^{t_f} \left(  
        (\check{\btheta}(\tau) - \btheta^*_f)^\top  \Q (\check{\btheta}(\tau) - \btheta^*_f)
 + R(\check{u}(\tau) - u^*_f)^2  \right) d\tau \\
& \text{subject to}
& & \ddot{\check{\theta}} = \frac{L}{I} u - \frac{\Chatd}{I} |\dot{\check{\theta}}| \dot{\check{\theta}} - \frac{\Lcm mg}{I} \sin(\check{\theta}), \\
& & & 0 = -\alpha\; \Phi + \left( \overline{\Delta}_{\dot\theta} + D + \eta \right),  \\
& & & 0 = - \lambda \Omega_{\text{max}} + \Phi,
        \quad \check{\btheta}(t_0) = \btheta^*_0, \\
& & & | \dot{\check{\theta}}(t)| \le \dot\theta_\text{max} - \dot{\tilde{\theta}}_\text{max},
        \quad | \check{u}(t)| \le u_\text{max} - \tilde{u}_\text{max},
        \quad  |\dot{\check{u}}(t)| \le \Delta u_\text{max}
\end{aligned}
\end{equation*}

where $\overline{\Delta}_{\dot\theta}$ is the maximum realization of $\Delta(\dot{\check{\theta}})$, and $\dot{\tilde{\theta}}_\text{max}$, $\tilde{u}_\text{max}$ are time-invariant and acquired from equations (\ref{eq:sc1}), (\ref{eq:sc2}), (\ref{eq:sc3}) by using $\Omega_{\text{max}}$, $\Phi = \lambda \Omega_{\text{max}}$, and $\alpha = ( \overline{\Delta}_{\dot\theta} + D + \eta ) / (\lambda \Omega_{\text{max}})$.
Since we are assuming the same physical system, physical parameters, constraints, and uncertainty parameters are shared between the controllers.

\section{Implementation}\label{sec:implementation}

\subsection{System Overview}
\begin{figure}[t]
    \centering
    \begin{subfigure}[b]{0.34\textwidth}
        \includegraphics[width=1\textwidth]{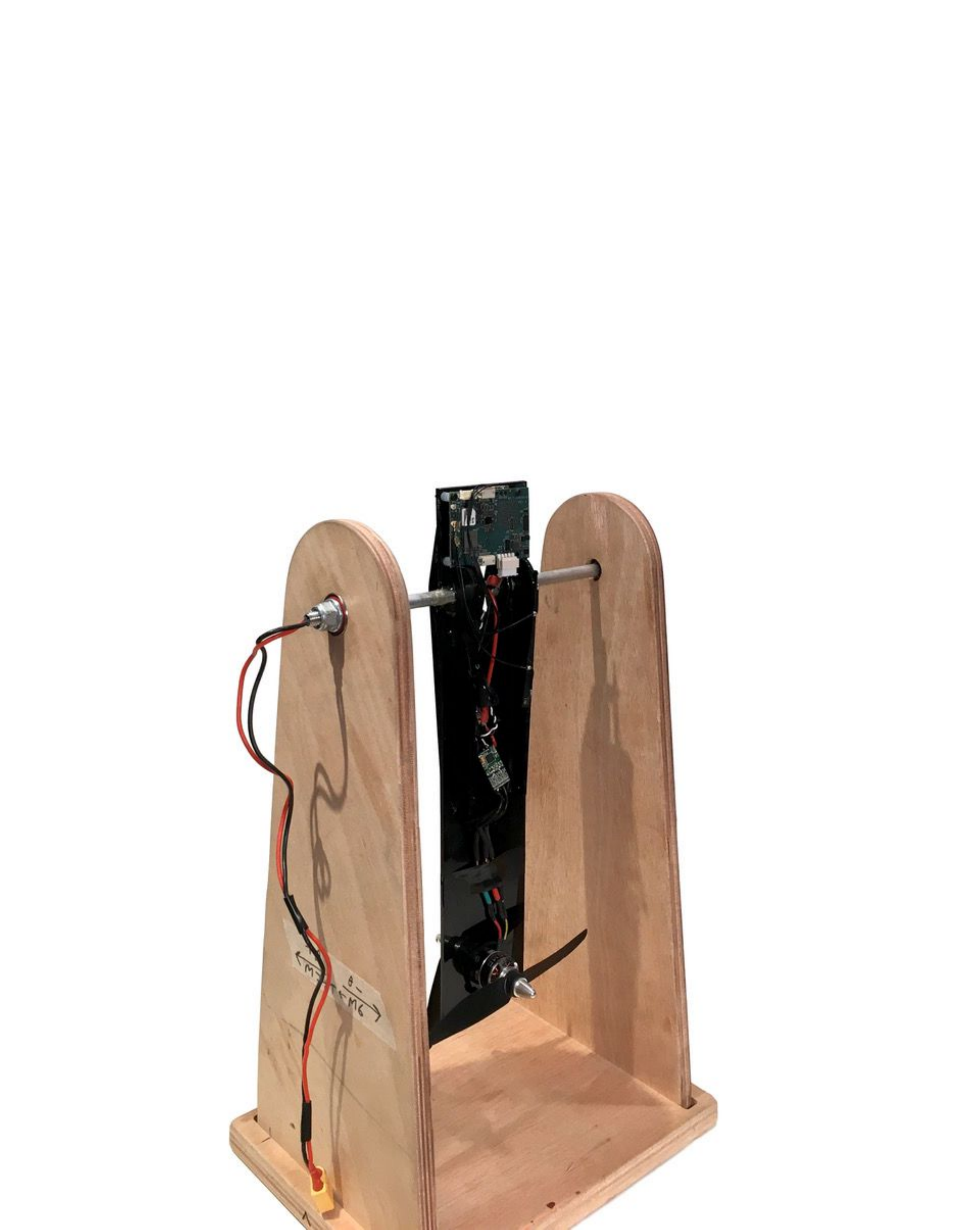}
        \caption{No attachments}
        \label{fig:pend-noattach}
    \end{subfigure}
    \begin{subfigure}[b]{0.32\textwidth}
        \includegraphics[width=1\textwidth]{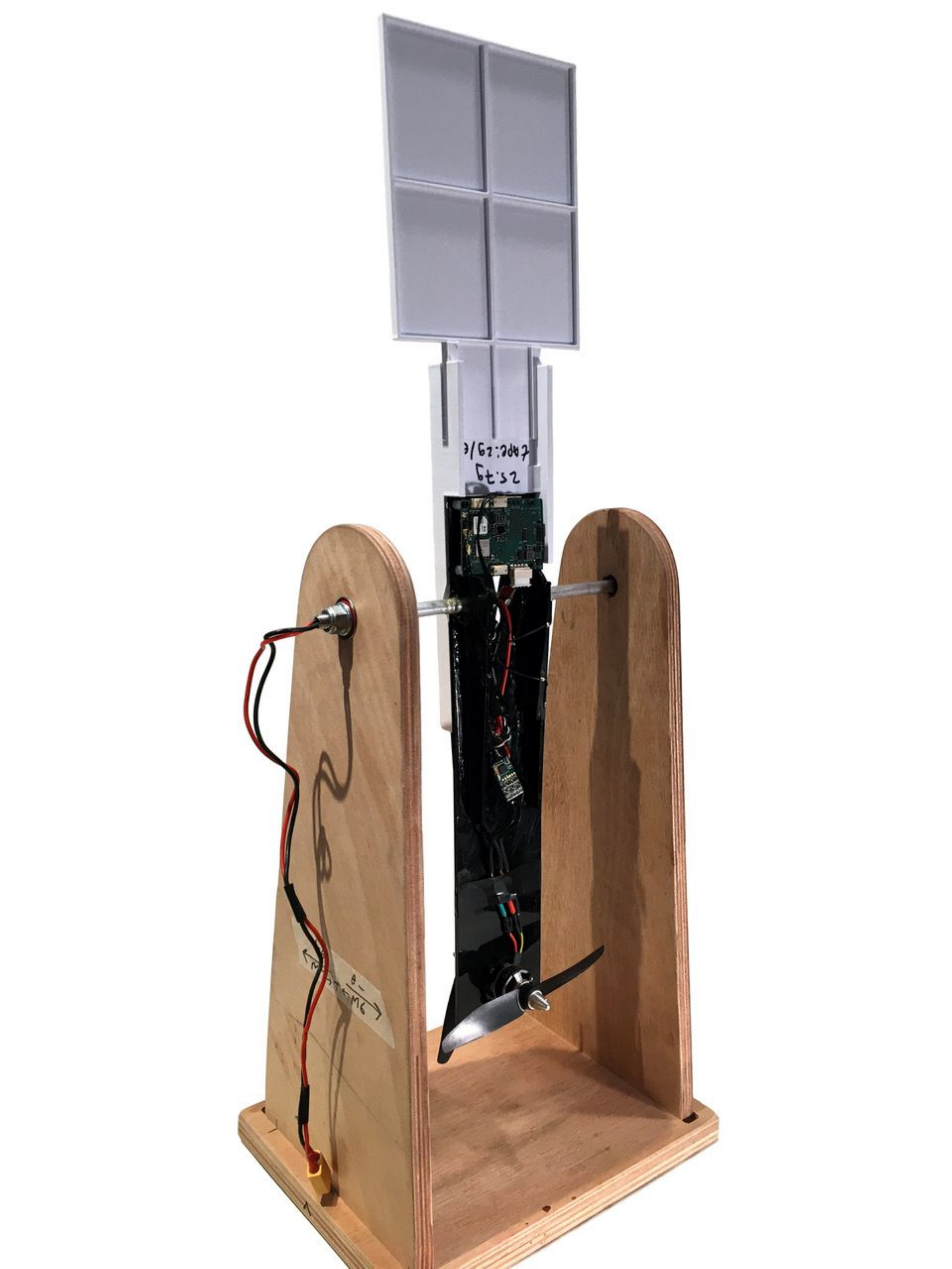}
        \caption{Flat Plate attachment}
        \label{fig:pend-noattach}
    \end{subfigure}
    \begin{subfigure}[b]{0.33\textwidth}
        \includegraphics[width=1\textwidth]{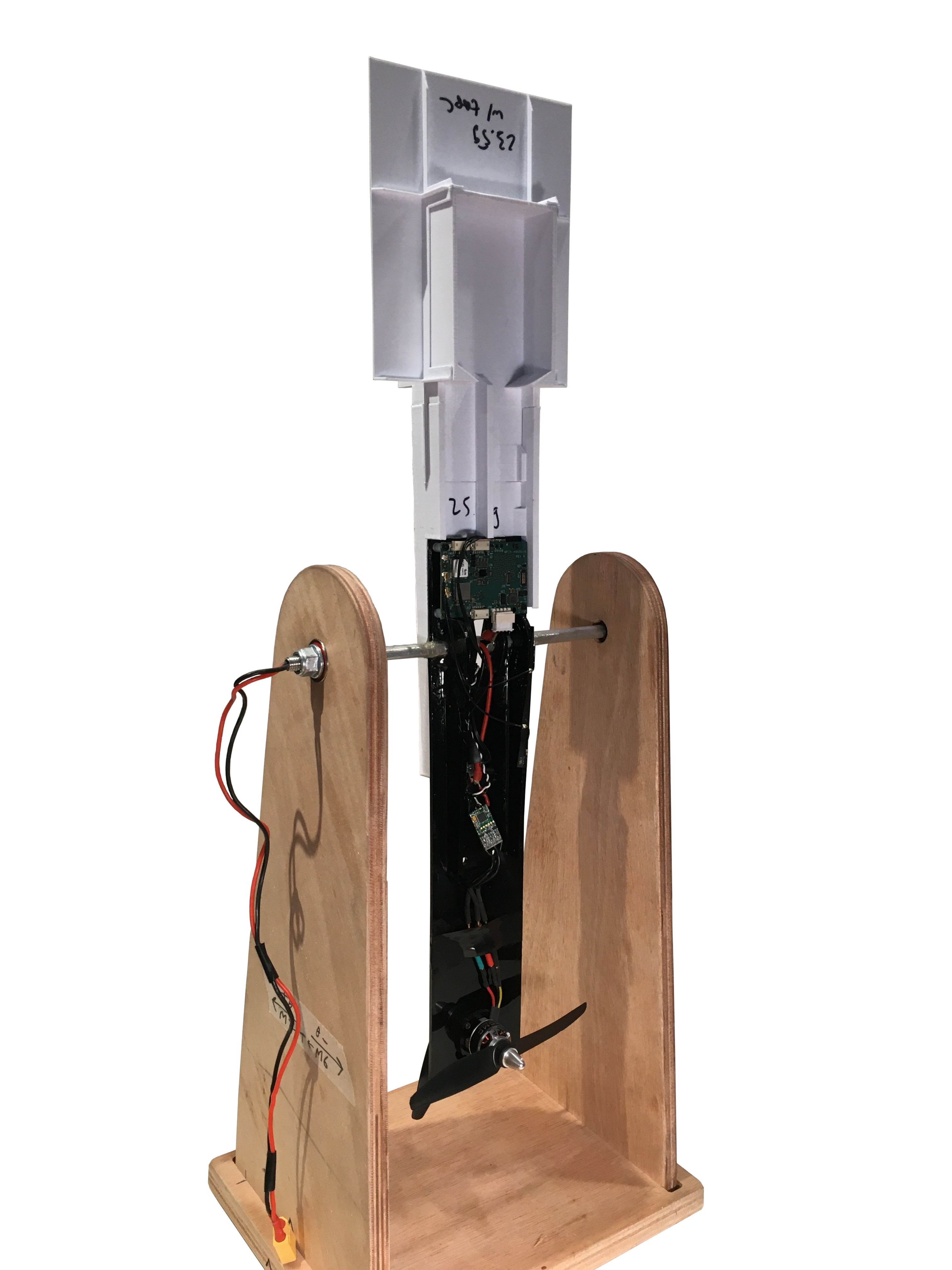}
        \caption{Scoop attachment}
        \label{fig:pend-noattach}
    \end{subfigure}
    \caption{
    TMPC, DTMPC, and ADTMPC are tested on a custom-built hardware pendulum. Different hardware configurations were designed, pictured in order of increasing drag.
    }
    \label{fig:pend}
\end{figure}

For the real-world comparison and performance analysis of the presented trajectory tracking algorithms, we designed a custom pendulum testbed, shown in Fig.~\ref{fig:pend}.
The pendulum arm swings freely around the shaft and is actuated with two propeller motors, mounted on either side of the arm.
Each propeller motor is controlled with an electronic speed controller that interfaces with an embedded computer using pulse-width modulation (PWM).
To accurately command desired thrusts, we use thrust-to-PWM lookup tables that were acquired from a dynamometer.
Using our CAD model, the physical parameters of the pendulum were estimated and are listed in Tables~\ref{tbl:parameters} and~\ref{tbl:hw_platforms}.
Our algorithms are implemented in C++ using ROS~\cite{quigley2009ros} and run in real-time on the Qualcomm Snapdragon Flight embedded computer attached to the pendulum arm.

As seen in Fig.~\ref{fig:pend}, three hardware configurations with different drag coefficients were built---this was done to show ADTMPC's ability to recover model parameters in different settings.
We used the pendulum with the flat plate attachment for all other experiments due to the fact that it has the highest drag to inertia ratio, making the nonlinearity due to drag more apparent.

\begin{table}[t!]
    \ra{1.2}
	\centering
	\caption{Experiment Parameters.}
	\begin{tabular}{@{}ll @{\hspace{4em}} ll @{\hspace{4em}} ll @{\hspace{4em}} ll@{}} %
	\toprule
	Param. & Value & Param. & Value & Param. & Value & Param. & Value \\ 
	\midrule

	$\Omega_{\text{max}}$ {\scriptsize (Sec.~\ref{sec:results-dtmpc-disturb})} & \SI{7.5}{\degree}
	& $N$             & 45
	& $Q$             & $\mathrm{diag}(10, 0.1)$
	& $R$             & 1.0
	\\
	$\Omega_{\text{max}}$ {\scriptsize (Sec.~\ref{sec:results-adtmpc})} & \SI{11.5}{\degree}
	& $\Delta t$      & \SI{0.010}{\second}
	& $Q_f$           & $\mathrm{diag}(10, 0.1)$
	& $M$             & 0.01
	\\
    $\hat{C}_d$ {\scriptsize (Sec.~\ref{sec:results-dtmpc})}  & \SI{1.0}{\gram\meter\squared}
    & $D$                      & \SI{20}{\per\square\second}
    & $\Delta\alpha_\text{max}$& \SI{300}{\per\square\second}
    & $u_\text{max}$           & \SI{2.0}{\newton}
	\\
	$\tilde{C}_\text{d}$ {\scriptsize (Sec.~\ref{sec:results-dtmpc})}      & \SI{1.0}{\gram\meter\squared}
	& $\overline{\Delta}_{\theta}$             & \SI{20}{\per\square\second}
	& $\Delta u_\text{max}$    & \SI{4.5}{\newton\per\second}
	&  $\dot{\theta}_\text{max}$    & \SI{15}{\radian\per\second}
	\\
	$\eta$                   & \SI{0.1}{\per\second\squared}
	& $\lambda$                & \SI{4.0}{\per\second}
	& $\alpha_\text{min}$     & \SI{40}{\per\second}
	& $\alpha_\text{max}$      & \SI{140}{\per\second}
	\\
	\bottomrule
	\\
	\end{tabular}
	\label{tbl:parameters}
\end{table}

\subsection{Technical Details}

For digital implementation, Problems 3.1 and 3.3 were discretized with a time horizon $N$ and a timestep~$\Delta t$.
Problems 3.1 and 3.3 are inherently non-convex due to the nonlinear pendulum dynamics and the non-convex DTMPC constraints.
Sequential convex programming, similar to that in~\cite{mao2018successive}, was used to create convex subproblems, which were then solved using CVXGEN at approximately 6-12 Hz~\cite{cvx,gb08}.
Problem 3.2 requires the solution of eight LPs as previously described, producing updated parameter bounds at a rate of \SI{50}{\Hz}.
In all experiments, the online BLSC controller runs at 500 Hz.

The SMID framework of Problem 3.2 requires state and control measurements.
We computed angular acceleration by numerical differentiation and carefully ensured that all signals are sampled at the same time instant (i.e., all signals are equally delayed to match the lag caused by numerical differentiation).
For the purpose of outlier rejection, parameters are only accepted after they are validated by two consecutive results.

\subsection{Experiment Description}
We conducted two sets of experiments.
First, in Section~\ref{sec:results-dtmpc} we compare DTMPC to TMPC and illustrate the ability of DTMPC to leverage state-dependent uncertainty.
Second, in Section~\ref{sec:results-adtmpc} we highlight the ability of ADTMPC to reduce uncertainty of model parameters, which leads to improved performance over DTMPC.

DTMPC experiments begin with the pendulum at rest at zero angle and end once the pendulum has reached steady state at the desired setpoint angle of $5\pi$.
As adaptation requires sufficient excitation of the dynamics, ADTMPC experiments use longer trajectories.
Starting at rest at zero angle, two desired setpoints are specified.
After reaching steady state at the first setpoint, the other setpoint is provided.
The pendulum cycles between steady state at these setpoints until the rate of adaptation is zero for 15 seconds.
Experimental parameters are listed in Table~\ref{tbl:parameters}.

\section{Experimental Results: DTMPC}\label{sec:results-dtmpc}
DTMPC leverages state-dependent constraints and uncertainty to perform aggressive and conservative behaviors only when circumstances demand it.
In this section we will experimentally compare the performance of DTMPC and TMPC controllers in changing environments, quantifying DTMPC's reduction of control effort and ability to maintain higher speeds.
We will first examine the case where the tube size is constrained in certain segments of the state-space, requiring low tracking error in particular regions.
We will then look at the case where certain segments of the state-space possess higher uncertainty, akin to specific locations in the environment having more disturbance and noise.

\subsection{DTMPC: Locally Restricted Tubes }\label{sec:results-dtmpc-tubes}
\begin{figure}[t]
    \centering
    \begin{subfigure}[b]{0.48\textwidth}
        \includegraphics[width=1\textwidth]{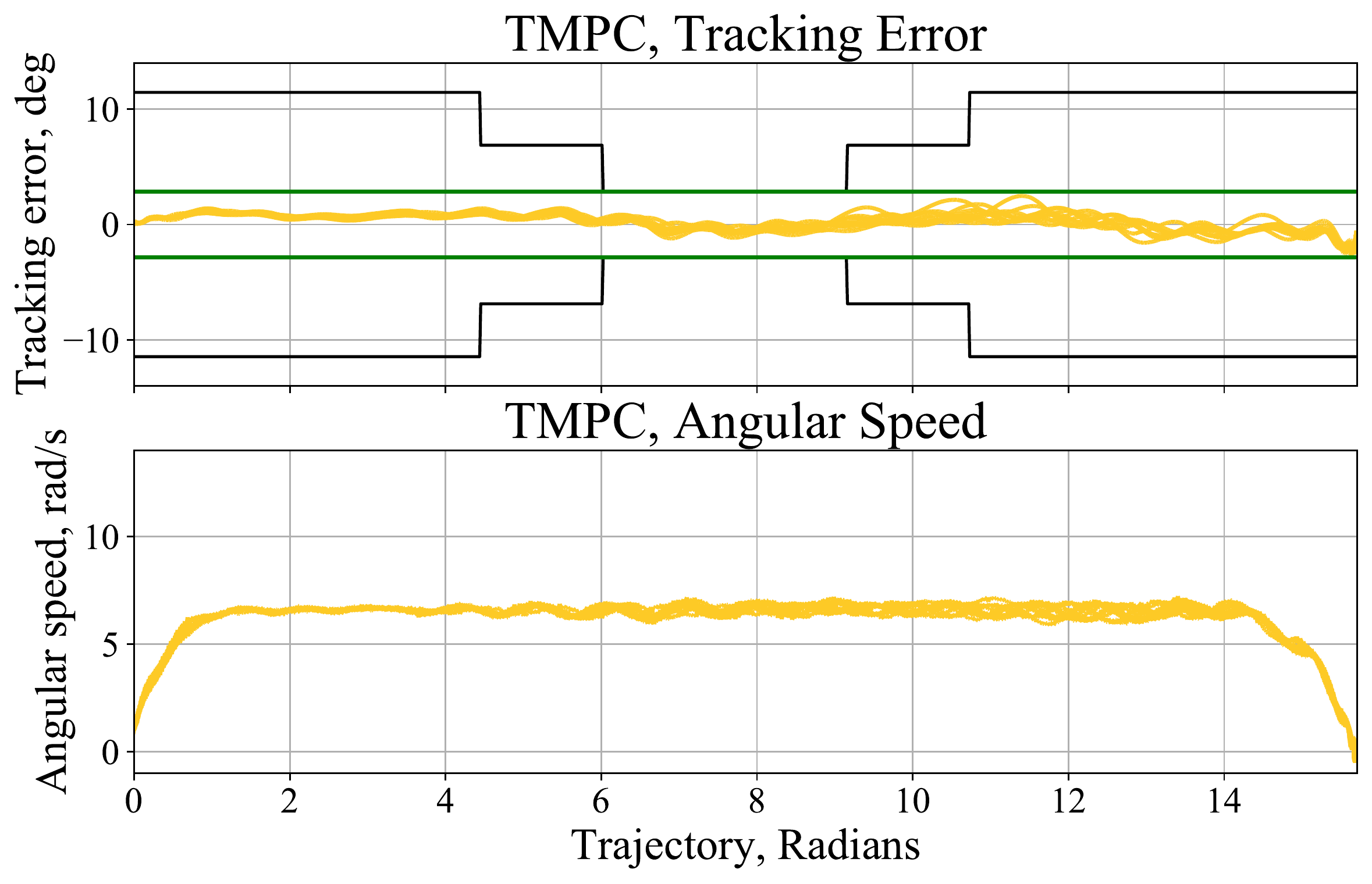}
        \caption{TMPC}
        \label{fig:tmpc-obst}
    \end{subfigure}
    \begin{subfigure}[b]{0.49\textwidth}
        \includegraphics[width=1\textwidth]{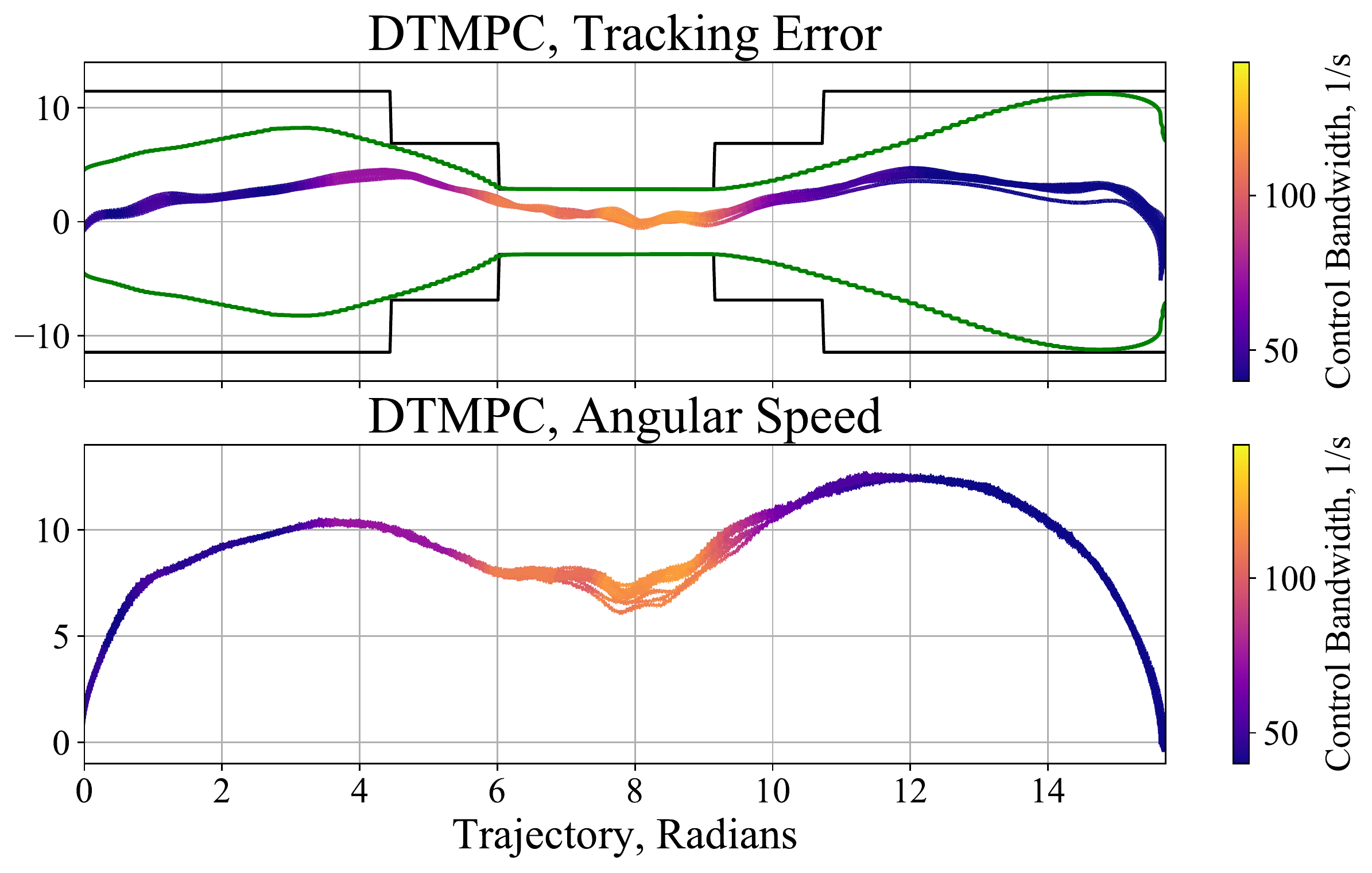}
        \caption{DTMPC}
        \label{fig:dtmpc-obst}
    \end{subfigure}
    \caption{
    Comparison of DTMPC and TMPC in a scenario with locally tight constraints (black).
    Tracking error (top) is colored according to control bandwidth.
    TMPC (left) generates a constant-sized RCI tube (green) around the trajectory (top left), significantly restricting its maximum speed (bottom left).
    DTMPC (right) varies the size of the RCI tube (green) to ensure that the system satisfies the constraints (black).
    Angular speed is reduced only in the vicinity of the restricted region (bottom right).
    Experiments repeated 10 times.
    }
    \label{fig:obst_avoid}
\end{figure}

We demonstrate that DTMPC renders an effective local response without sacrificing overall performance, even in the presence of locally tight tube constraints.
Tracking error constraints $\Omega_{\max}(\theta)$ are defined throughout the trajectory as
\begin{gather*}
\Omega(\theta) \le \Omega_{\max}(\theta) = 
\left\{
\begin{array}{ll}
\SI{0.05}{\radian} & \text{if} ~~ 2\pi \le \theta \le 3\pi \\
\SI{0.1}{\radian} & \text{if}  ~~ 1.5\pi \le \theta \le 2\pi ~~ \text{or} ~~ 3\pi \le \theta \le 3.5\pi \\
\SI{0.2}{\radian} & \text{else}  \\
\end{array}
\right.
\end{gather*}
and are depicted in Fig.~\ref{fig:obst_avoid} as thick black lines.

High uncertainty and small tube size require TMPC to utilize extremely high bandwidth to ensure theoretical robustness.
However, aggressive solutions with such high bandwidth are not realizable due to actuator limitations, as the motors are rate-limited.
Hence the controller's bandwidth cannot be arbitrarily high and must be no larger than $\alpha_{\text{max}}$.
For the system to use $\alpha_{\text{max}}$ and still remain within the tight tube around the trajectory, the anticipated uncertainty must be reduced.
This forces TMPC to use slower speeds to complete the task, as depicted in Fig.~\ref{fig:tmpc-obst}.

Unlike TMPC, DTMPC is only conservative when it has to be---in the vicinity of the constrained region.
To accommodate for a tighter constraint, DTMPC only temporarily reduces its speed from 10 to \SI{7}{\radian\per\second}, while also increasing its control bandwidth (represented by a warmer signal color).
When the constraint is no longer present, DTMPC allows for a larger tracking error in the controller, increasing the size of the RCI tube to match new constraints---from \SI{0.05}{\radian} up to \SI{0.2}{\radian}, more than four-fold.
The increase in available tube size allows DTMPC to exhibit less conservative behaviors, as seen by an increase in angular speed in Fig.~\ref{fig:dtmpc-obst}.

Quantitative results of these experiments are presented in the top half of Table~\ref{tbl:tmpc-dtmpc}.
Compared to TMPC, DTMPC produces overall less conservative trajectories, taking 31\% less time to get to the goal, while using significantly less control effort (34\% reduction) to achieve that.
This experiment clearly depicts DTMPC's ability to leverage state-dependent constraints.

\begin{table}[t]
    \ra{1.2}
	\centering
	\caption{Quantitative Analysis of the Conservativeness and Control Effort of DTMPC and TMPC controllers.}
	\begin{tabular}{@{}llrrrr@{}} %
	\toprule
	  & Metric          &  TMPC       & DTMPC      & Reduction & Increase \\ 
	\midrule
	                              & Control Effort &  \SI{1.492}{\newton\squared\second}  & \SI{0.99}{\newton\squared\second} &  33.9 \%& --- \\
	A: Locally Restricted Tubes  & Rise Time          &   1.90 s               & 1.31 s          & 31.0 \% & ---  \\
	(averaged over 10 trials)& Maximum Speed & \SI{7.1}{\radian\per\second} & \SI{12.62}{\radian\per\second} &  --- & 178 \%  \\
	& Mean Tracking Error & \SI{0.77}{\degree}  & \SI{1.78}{\degree}  & --- &  233 \%  \\
	\midrule
	                              & Control Effort &  \SI{1.96}{\newton\squared\second}  & \SI{1.34}{\newton\squared\second} &  31.8 \%& --- \\
	B: Locally Increased Disturbance  & Rise Time          &   1.57 s               & 1.41 s          & 10.1 \% & ---  \\
	(averaged over 20 trials) & Maximum Speed & \SI{8.52}{\radian\per\second} & \SI{11.4}{\radian\per\second} &  --- & 133 \%  \\
	& Mean Tracking Error & \SI{1.15}{\degree}  & \SI{1.70}{\degree} & --- &  148 \%  \\

	\bottomrule
	\end{tabular}
	\label{tbl:tmpc-dtmpc}
\end{table}

\begin{figure}[t]
    \centering
    \begin{subfigure}[b]{0.48\textwidth}
        \includegraphics[width=1\textwidth]{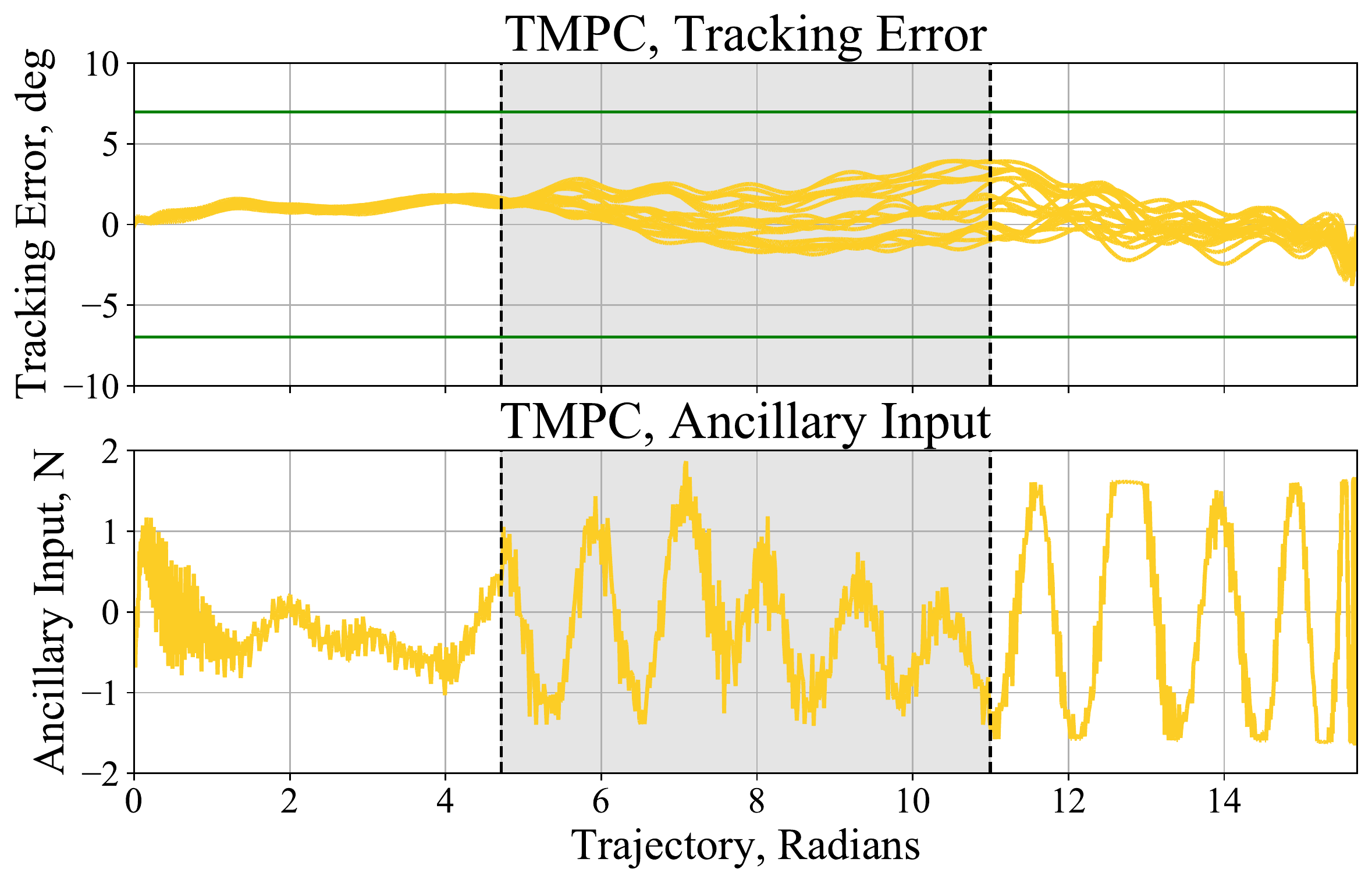}
        \caption{TMPC}
        \label{fig:tmpc-disturb}
    \end{subfigure}
    \begin{subfigure}[b]{0.49\textwidth}
        \includegraphics[width=1\textwidth]{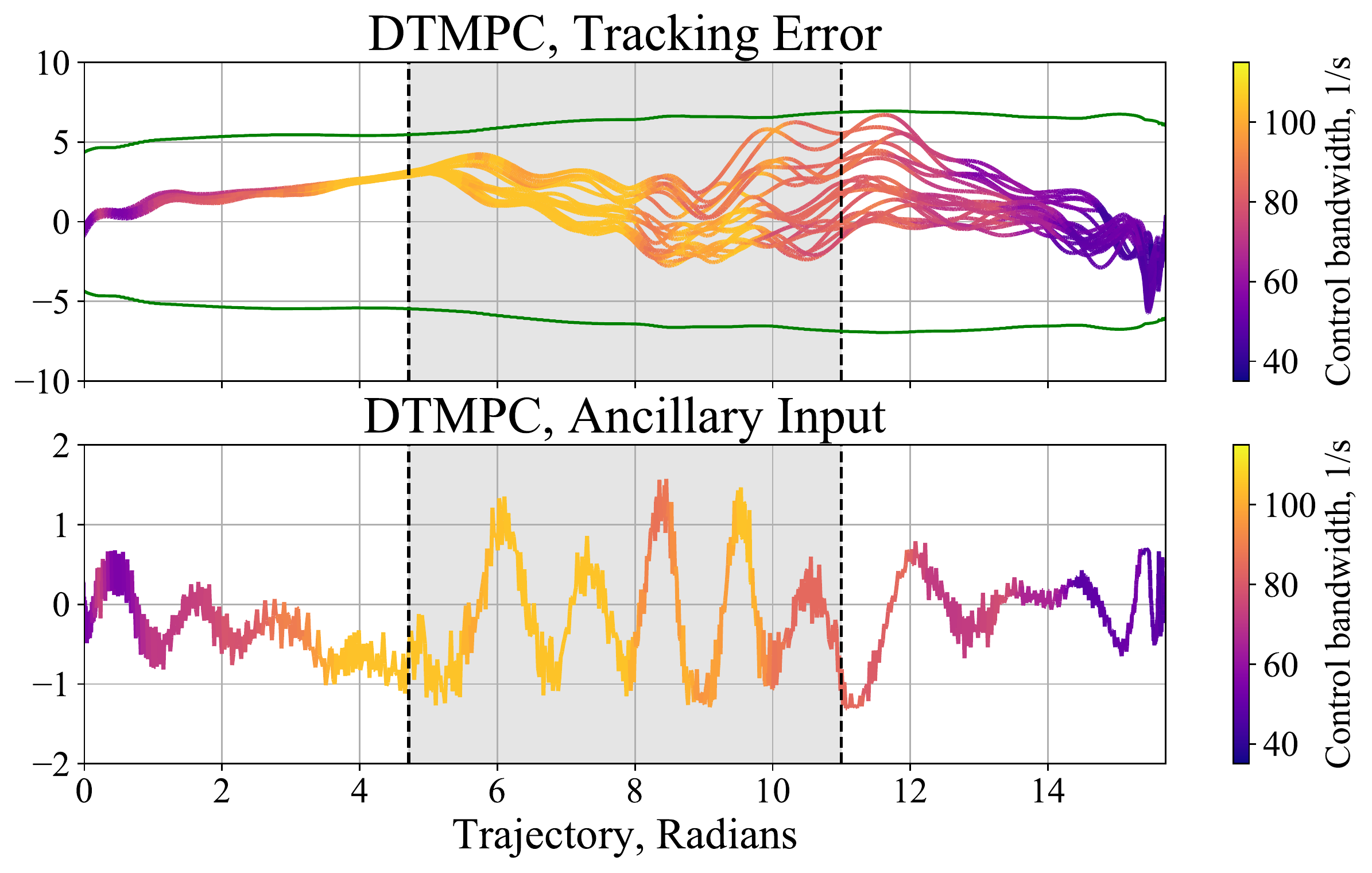}
        \caption{DTMPC}
        \label{fig:dtmpc-disturb}
    \end{subfigure}
    \caption{
    Comparison of DTMPC and TMPC in a scenario with locally increased background disturbance.
    Within the region of increased uncertainty (marked with shaded boxes), both policies utilize high-amplitude high-bandwidth control.
    After the disturbance, DTMPC gradually stabilizes to a desired trajectory without using as much effort---the uncertainty does not require it.
    In comparison, TMPC continues to use high control input, overcompensating and destabilizing the system.
    Experiments repeated 20 times.
    }
    \label{fig:disturb}
\end{figure}

\subsection{DTMPC: Locally Increased Disturbance} \label{sec:results-dtmpc-disturb}
Here, we examine the case where additional modeled uncertainty $\Delta(\theta)$ is introduced into the system. 
This uncertainty is assumed to be a function of state and is defined by
\begin{gather*}
\Delta(\theta) = 
\left\{
\begin{array}{ll}
r \,\overline{\Delta}_{\theta} & \text{if} ~~ 2\pi \le \theta \le 3\pi \\
\frac{1}{2} \, r \, \overline{\Delta}_{\theta} & \text{if}  ~~ 1.5\pi \le \theta \le 2\pi ~~ \text{or} ~~ 3\pi \le \theta \le 3.5\pi \\
0 & \text{else}  \\
\end{array}
\right.
\end{gather*}
where $\overline{\Delta}_{\theta}$ is the maximum observed realization of $\Delta(\theta)$ and $r$ is sampled from a uniform random distribution between $-1$ and $1$. 
The realization of the uncertainty $\Delta(\theta)$ is implemented as an additional acceleration term added to the control input at every time step.
Since $\Delta(\theta)$ is state-dependent uncertainty, TMPC has no way of handling it but to assume its largest anticipated realization.
In contrast, DTMPC can leverage the state-dependent nature of this uncertainty.

Results are presented in Fig.~\ref{fig:disturb} and the second half of Table~\ref{tbl:tmpc-dtmpc}.
As the system gets closer to the uncertain region (marked as the shaded boxes), DTMPC increases its control bandwidth and uses high-amplitude high-bandwidth control to counteract random motor disturbances.
After passing through this region, DTMPC decreases its control bandwidth and utilizes less feedback effort to stabilize the system to the desired trajectory.
In contrast, TMPC is destabilized by the uncertain region: it continues to use excessive feedback, overcompensating for any small perturbations.
TMPC ends up producing an oscillatory response at a naturally unstable steady-state point ($5\pi$), a result of TMPC's uniform usage of a high-bandwidth high-effort policy.
This is a downside of using a fixed one-size-fits-all control policy: to ensure robustness in high-uncertainty regions, TMPC may inadvertently sacrifice stability in low-uncertainty segments.

As a consequence of such behavior, TMPC renders unnecessarily accurate tracking along the entire trajectory.
Although TMPC's average tracking error is 30\% smaller than that of DTMPC, TMPC used on average 46\% more control effort than DTMPC to accomplish this.
We stress that this tracking improvement is unnecessary, bearing in mind that the system is not close to exceeding its constraints.
Indeed, DTMPC and TMPC require the system to stay within their respective tubes, not to perform perfect trajectory tracking.
This experiment shows DTMPC's ability to leverage state dependent uncertainty by utilizing aggressive control effort only when necessary.

\section{Experimental Results: ADTMPC}\label{sec:results-adtmpc}

ADTMPC steadily decreases parameter uncertainty, leading to less conservative trajectories, reduced tracking error, and reduced control effort.
In this section we aim to illustrate these behaviors.
We will first indicate the advantages of online adaptation by comparing ADTMPC with DTMPC. 
We will then show that our adaptation framework is generalizable by demonstrating its ability to produce consistent results for various controllers, trajectories, and hardware configurations.

\begin{figure}
    \centering
    \includegraphics[width=1\textwidth]{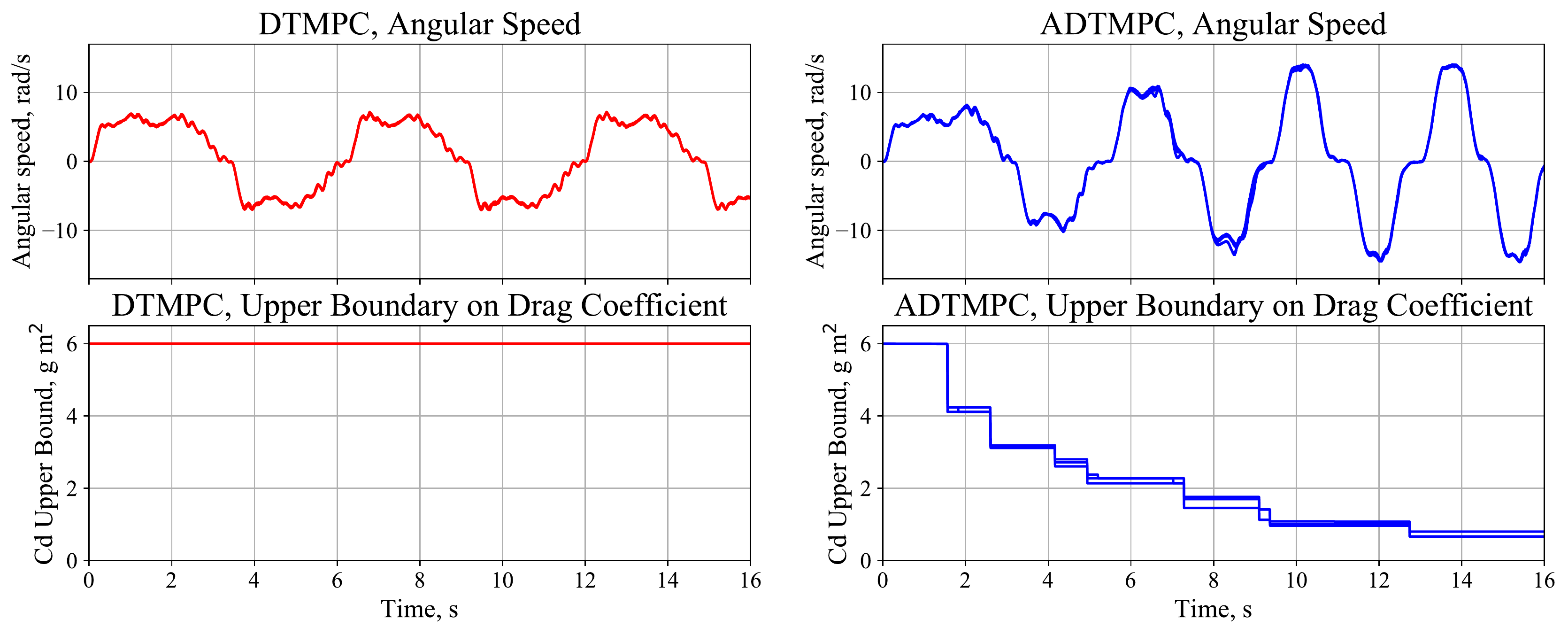}
    \caption{
    Comparison of DTMPC and ADTMPC control policies: angular speed and the upper bound on $C_d$. 
    With every cycle, ADTMPC reaches higher speeds due to the decrease in $\overline{C_d}$, until reaching steady-state.
    DTMPC performance is provided as the baseline to ADTMPC. 
    Experiment repeated 3 times.
    }
    \label{fig:adtmpc_dtmpc}
\end{figure}

\subsection{ADTMPC and DTMPC: Performance Comparison}
We first illustrate the potential of adaptation when applied online to an uncertain system.
In these experiments we start the DTMPC and ADTMPC controllers at an initial zero angle condition with identical initial boundaries on the coefficient of drag.
We then run the controllers in a cyclic trajectory between the angles of $5\pi$ and $\pi$.

Fig.~\ref{fig:adtmpc_dtmpc} depicts the results of these experiments. 
In each cycle, ADTMPC improves its model estimates and reduces parameter uncertainty.
Reduced uncertainty allows the system to achieve higher speeds with every cycle while remaining within the desired tube.
After a certain point, ADTMPC no longer excites the unmodeled dynamics enough to further reduce the uncertainty, reaching a steady-state condition for the system. 
On average, we observed an overall uncertainty reduction of 89\% from the original uncertainty.

Table~\ref{tbl:adtmpc_quantitative} presents the quantitative evaluation of the ADTMPC performance during the adaptation stage. 
As parameter uncertainty is reduced, the controller generates more accurate trajectories, better matching the system's actual behavior.
As a result of adaptation, tracking error is reduced by 34\%, despite using 35\% less ancillary BLSC effort to maintain the system in the RCI tube.
Reduced uncertainty also allows ADTMPC to generate less conservative trajectories: throughout adaptation, the maximum speed realized in the trajectory is increased by 80\%, from an average of \SI{8}{\radian\per\second} up to \SI{14.5}{\radian\per\second}. 
This increase allows ADTMPC to complete the cycle trajectory 1.8 times faster than DTMPC.

\begin{table}[t]
    \ra{1.2}
	\centering
	\caption{Quantitative Analysis of ADTMPC: online adaptation steadily reduces tracking error and ancillary control effort, increases speed. DTMPC provided as the baseline.}
	\begin{tabular}{@{}llrrr@{}} %
	\toprule

	& $\overline{C_d}$, Drag Upper Bound  & 4.0-6.0 \SI{}{\gram\meter\squared} & 2.0-4.0 \SI{}{\gram\meter\squared} & 0.0-2.0 \SI{}{\gram\meter\squared} \\
	\midrule
	& Average Ancillary Input & \SI{0.421}{\newton} & \SI{0.403}{\newton} & \SI{0.279}{\newton} \\
	ADTMPC & Average Tracking Error & \SI{2.761}{\degree} & \SI{2.167}{\degree} & \SI{1.808}{\degree} \\
	& Average Maximum Speed & \SI{8.11}{\radian\per\second} & \SI{10.82}{\radian\per\second} & \SI{14.59}{\radian\per\second} \\
	
	\midrule
	& Average Ancillary Input & \SI{0.428}{\newton} & \SI{0.417}{\newton} & \SI{0.304}{\newton} \\
	DTMPC & Average Tracking Error & \SI{2.513}{\degree} & \SI{2.371}{\degree} & \SI{1.881}{\degree} \\
	& Average Maximum Speed & \SI{7.99}{\radian\per\second} & \SI{9.12}{\radian\per\second} & \SI{13.09}{\radian\per\second} \\
	
	\bottomrule
	\end{tabular}
	\label{tbl:adtmpc_quantitative}
\end{table}

\subsection{ADTMPC: Consistency Over Various Trajectories and Controllers}
In this section we illustrate that ADTMPC is generalizable to a variety of trajectories and ancillary controllers.
We show that the adaptation framework consistently converges to similar values of the drag coefficient when applied to different cyclic trajectories, as well as when varying the time constant $\lambda$ in the BLSC controllers.

\begin{figure}
    \centering
    \includegraphics[width=1\textwidth]{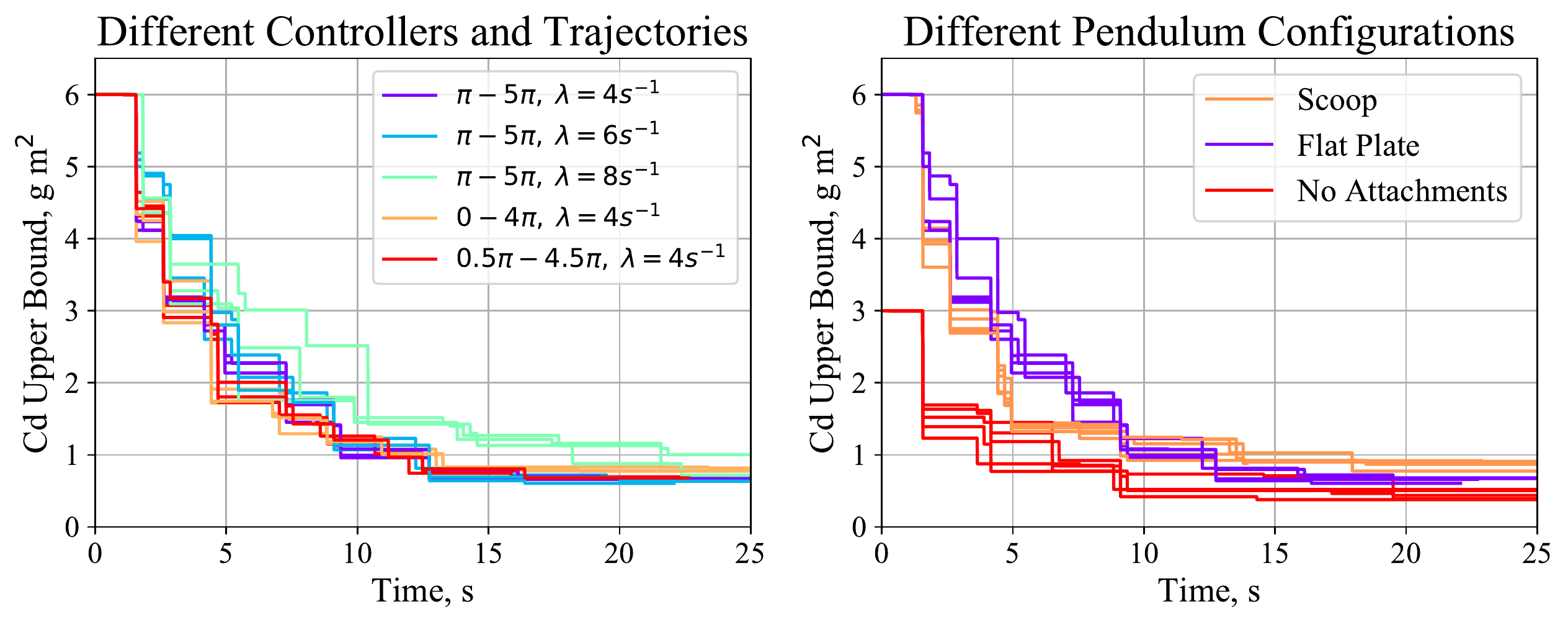}
    \caption{
    ADTMPC produces consistent reductions in uncertainty over a variety of cyclic trajectories and BLSC controllers (left).
    ADTMPC is generalizable to various hardware configurations (right).
    Notably, as seen around \SI{25}{\second}, larger reductions are observed for systems with lower expected drag.
    Each experiment repeated 5 times.
    }
    \label{fig:adtmpc_diff}
\end{figure}

The left side of Fig.~\ref{fig:adtmpc_diff} depicts the results of these experiments.
We note that certain trajectories excite the unmodeled dynamics less than others, producing slightly more conservative estimates.
We also point out that more aggressive controllers are faster at cancelling disturbances and hence do not excite the unmodeled dynamics as much, producing slightly more conservative values.

\subsection{ADTMPC: Consistency Over Various Hardware Configurations}
In these experiments we aim to show that ADTMPC is generalizable to a variety of hardware configurations.
Hardware parameters of the pendulum were varied through 3D-printed attachments, pictured in Fig.~\ref{fig:pend}.
The motivation behind these attachments was to change the drag coefficient of the system (and consequently the mass, inertia, and center of mass).
Parameters for these hardware configurations are presented in Table~\ref{tbl:hw_platforms}.

As expected, we observed ADTMPC to consistently reduce drag coefficient uncertainty and repeatedly converge to the appropriate steady-state condition for each hardware configuration.
Results of online adaptation are presented in Fig.~\ref{fig:adtmpc_diff}, and average steady-state conditions are given in Table~\ref{tbl:hw_platforms}.
We note a favorable behavior in the results of adaptation between the hardware configurations: final boundaries on the coefficient of drag are proportional to the expected amount of drag generated by each pendulum.
This illustrates that ADTMPC performs uniformly well for a variety of physical configurations of a given hardware system.

\begin{table}[]
    \ra{1.2}
	\centering
	\caption{Parameters for Varied Pendulum Hardware Configurations. }
\begin{tabular}{@{}l|lllll|l@{}}
                           & \multicolumn{4}{l}{Parameter Values Acquired From CAD}                  & ADTMPC Results                  \\ \midrule
Pendulum Configuration     & CoM, $L_{cm}$ & Mass, $m$ & Inertia, $I$ & $L$  & Theoretical $C_d$ & Upper Boundary, $\overline{C_d}$ \\ \midrule

No Attachments & \SI{9.21}{\cm}                  & \SI{218}{\g}     & \SI{4.2}{\g\meter\squared}       & \SI{23}{\cm} & \SI{0.204}{\g\meter\squared}          & \SI{0.43}{\g\meter\squared}                         \\
With the Flat Plate     & \SI{2.17}{\cm}                  & \SI{309}{\g}     & \SI{6.9}{\g\meter\squared}       & \SI{23}{\cm} & \SI{0.46}{\g\meter\squared}           & \SI{0.65}{\g\meter\squared}                         \\
With the Scoop      & \SI{0.49}{\cm}                   & \SI{343}{\g}     & \SI{7.6}{\g\meter\squared}       & \SI{23}{\cm} & \SI{0.55}{\g\meter\squared}           & \SI{0.86}{\g\meter\squared}                         \\ \bottomrule
\end{tabular}
\label{tbl:hw_platforms}
\end{table}

\section{Discussion}\label{sec:discussion}

\subsection{Dynamic Tube MPC}

The biggest advantage of DTMPC over TMPC is its ability to dynamically adjust to changing environments. 
DTMPC only applies aggressive stabilizing control effort when the constraints and uncertainty demand it, reducing the overall amount of expanded control effort by over 30\%.
This makes DTMPC a more efficient controller that does not take needless actions.
As seen from the trajectory tracking error graphs in Fig.~\ref{fig:dtmpc-obst} and~\ref{fig:dtmpc-disturb}, DTMPC prefers less aggressive behaviors unless high uncertainty or proximity to constraint boundaries require it to respond.
DTMPC also utilizes state-dependent constraints to avoid exhibiting uniformly conservative behaviors, as a result allowing for up to 80\% larger speeds to be achieved.
Although these improvements come at the cost of increased computational complexity, they are necessary to handle tasks with changing settings and objectives and do not prohibit DTMPC from running real-time.

These features of DTMPC make it a favorable choice for applications.
For instance, DTMPC could be applicable to spacecraft maneuvering and rendezvous, where its ability to minimize control effort and control bandwidth may extend equipment lifespan in orbit.
Additionally, we see DTMPC's potential in UAV collision avoidance.
DTMPC would dynamically adjust its control policy to become more aggressive in close proximity to obstacles while maintaining a non-conservative control policy otherwise.
This is supported by the results in Section~\ref{sec:results-dtmpc-tubes}, which presents DTMPC's performance in locally limited tube size environments.

Due to the nature of the DTMPC algorithm, this control policy reaches its peak performance when multiple sources of variable uncertainty exist---in other words, if state-dependent uncertainty is present.
If state-dependent uncertainty is marginal or nonexistent, DTMPC will possess similar performance characteristics to TMPC.

\subsection{Adaptive Dynamic Tube MPC}

We have shown ADTMPC to be reliable and effective in a variety of environments, control settings, and hardware configurations. 
The key features of the algorithm are that it preserves recursive feasibility throughout adaptation, which ensures consistently robust performance, and the auxiliary nature of SMID, which allows it to be executed in parallel with the main DTMPC framework at any convenient rate.

This makes ADTMPC advantageous in the scenarios when uncertain systems experience physical changes.
For instance, this policy could be leveraged in the field of package delivery, where an unknown cargo is added to aerial or ground robots, changing vehicle's physical properties (mass, CoM, inertia), but not the dynamics.
In this scenario, quick online short-term adaptation of the control and trajectory policies would allow reliable control of the modified system while maintaining robustness.
This approach alleviates the need for configuration-specific control policies, precise cargo positioning, and parameter tuning, creating a more versatile system.

We emphasize the difference between our approach to adaptation such as the L1 or Model Reference Adaptive Control (MRAC) frameworks.
Adaptation in these policies is used to adjust the parameters in the control law to adapt to current conditions.
In comparison, our adaptation framework is aimed at reducing the uncertainty of the overall model, as a result reducing the tracking error.
Our goal is to make overall conclusions about the system, not to improve its instantaneous performance.

Due to the nature of ADTMPC, we note that a tight boundary estimate on unmodeled uncertainty is desirable in order to achieve better adaptation results.
At the same time, such boundary must be an overestimate of actually experienced uncertainty---if disturbance excitation exceeds the given $D$, adaptation would converge to a wrong value.

\section{Conclusion}\label{sec:conclusion}

We showed that DTMPC outperforms its robust TMPC predecessor.
When applied to a nonlinear pendulum testbed, DTMPC designed uncertainty-aware trajectories that are faster, less conservative, and more energy-efficient.
Leveraging state-dependent constraints and uncertainty, it is the DTMPC's coupled tube-trajectory design that leads to the algorithm's versatility.
We see DTMPC as being particularly beneficial in changing objectives and environments.

Further, we evaluated Adaptive DTPMC -- an extension of DTMPC that uses SMID as the adaptation framework to improve its prediction model.
When applied online, the algorithm refines model parameters, generating more accurate and less conservative trajectories.
Our results show that ADTMPC is a flexible framework that can be applied to different instances of the same system without retuning the controller.
We believe this makes ADTMPC a less conservative, more efficient, and more accurate robust controller choice, compared to TMPC.

DTMPC and ADTMPC are unique among other tube MPC methods, and in this paper we have validated these algorithms, emphasizing their ability to successfully and efficiently operate in different environments. 
In the future, we plan to apply DTMPC and ADTMPC to multirotor vehicles, leveraging the ability of these algorithms to balance high-effort, aggressive control during obstacle avoidance together with low-effort, smooth control during nominal flight.

\bibliography{main.bbl}

\end{document}